# Is the existence of unbounded operators a problem for quantum mechanics?
## In response to Carcassi, Calderón, and Aidala

*Zhonghao Lu*
*Department of History and Philosophy of Science, University of Pittsburgh*
*lu_zhonghao@pit.edu*
*https://orcid.org/0000-0002-7564-2652*

**Abstract**

In this paper I argue against Carcassi, Calderón, and Aidala's recent claim that the Hilbert spaces are unphysical and should be replaced with the Schwartz spaces in quantum mechanics, since Hilbert spaces include states with infinite expectation values for certain observables. I also review and discuss issues regarding unbounded operators in quantum mechanics raised by Streater and Wightman, Heathcote, and Lemos. I argue that the existence of infinite expectation values does not cause problems in quantum mechanics. On the other hand, replacing the Hilbert spaces with the Schwartz spaces would cause more issues, as it would exclude a class of meaningful Hamiltonian evolutions. I also discuss the question in literature whether reformulating quantum mechanics with essentially self-adjoint operators instead of self-adjoint operators may cause problems. I further analyse the hierarchies of the notions of "physicality" and possibility in fundamental physics, and suggest that "physicality" is a vague concept. Finally, I connect the problem raised by Carcassi, Calderón, and Aidala with the problem of the Hadamard condition in quantum field theory.

## 1. The unphysicality of Hilbert spaces?

In a recent paper "The unphysicality of Hilbert spaces", Carcassi, Calderón, and Aidala [1] argue that, in quantum mechanics, the conventional Hilbert space structure is too large to be physical. They have examined several features of the Hilbert space structure, including the inner product structure and completeness, and conclude that "the inner product structure is required to capture a necessary physical requirement" (p.14), whereas on the other hand the requirement of completeness is unphysical. More specifically, they consider a Cauchy sequence of states $\{\psi_i\}$ in the Hilbert space $L^2(\mathbb{R})$ with finite expectation values $<X>_{\psi_i}$ of an unbounded operator $X$. Further suppose that the limit of $<X>_{\psi_i}$ is $\infty$ when $i \to \infty$. Due to the completeness of the Hilbert space, this sequence converges to an element $\psi$ in $L^2(\mathbb{R})$, which however lies outside the domain of $X$. Carcassi, Calderón, and Aidala conclude that such a state is unphysical, at least when $X$ is the position operator. They claim that a diverging position expectation value "clearly" "does not make physical sense" (p.13). To avoid such difficulty, they tentatively suggest to replace the Hilbert space with the Schwartz space in the formalism of quantum mechanics. Their proposal has received criticism and debates [2, 11].

From their argument we can see that the completeness of the Hilbert space structure itself should not be blamed. Finite-dimensional Hilbert spaces are complete, but they admit no unbounded operators and do not lead to the difficulties considered by Carcassi, Calderón, and Aidala. On the other hand, an incomplete subspace of an infinite-dimensional Hilbert space may still admit states outside the domain of some unbounded operators, and therefore cannot avoid the problem of admitting "unphysical states".

The real problem is brought by the existence of unbounded operators in the formalism of quantum mechanics. Many meaningful physical observables in quantum mechanics, including positions, momenta, and many Hamiltonians, are represented by unbounded self-adjoint operators, which can only be defined on a dense proper subspace $\text{Dom}(X)$ of their corresponding Hilbert spaces. Strictly speaking, these operators lack definitions on states outside their domains. Consequently, Carcassi, Calderón, and Aidala regard these states as "unphysical" and should be removed from the formalism of quantum mechanics.

It is worth noting that the same issue was raised earlier by Adrian Heathcote in his 1990 paper "Unbounded Operators and the Incompleteness of Quantum Mechanics" published in *Philosophy of*

*Science*. However, Heathcote's work has not received much attention. Heathcote questions a most fundamental postulate of quantum mechanics that

> There is a one-to-one correspondence between the possible states of a system and the normed rays of a Hilbert space $\mathcal{H}$. [10, p.526]

Besides the problem of infinite expectation values of common physical observables as also discussed by Carcassi, Calderón, and Aidala, Heathcote raises another problem: if the state $\psi$ is outside the domain Dom($\boldsymbol{H}$) of the Hamiltonian operator $\boldsymbol{H}$ of the system, the Schrödinger equation

$$i d\psi/dt = \boldsymbol{H}\psi$$

cannot be applied. Heathcote makes a further bold claim that quantum mechanics "could be said to be incomplete in the sense that we do not have within the theory an algorithm for generating the appropriate restrictions on $\mathcal{H}$ (p.533)". Heathcote is unable to provide such restrictions, facing a considerable difficulty that "the domain restrictions on the unbounded operators that one is interested in will not coincide-simply because the domains do not coincide (p.529)". Moreover, Heathcote worries that the operators on the resulting restricted space are no longer self-adjoint, which would further undermine the fruitful formalism of quantum mechanics. Carcassi, Calderón, and Aidala provide an answer to Heathcote's task: the restricted space should be the Schwartz space.

In the following sections, I argue that, first, the existence of unbounded operators does not lead to any conceptual or physical problems as they worry about, and, second, the Schwartz space formalism suggested by Carcassi, Calderón, and Aidala can bring additional difficulties.

## 2. The existence of unbounded operators is not a problem

So far, I have discussed two difficulties brought by the existence of unbounded operators in quantum mechanics. First, the states with infinite expectation values of physical observables seem unphysical. Second, the Schrödinger equation cannot be applied to states outside the domain of the Hamiltonian operator of the system.

The second problem is easy to resolve. When **H** is an unbounded self-adjoint operator, though it is not defined on the whole Hilbert space, its spectral decomposition's projection-valued measure spans the entire space. Accordingly, **H**'s spectral decomposition yields a one-parameter unitary group a one-parameter group of unitary operators exp(-i**H**t), which are bounded operators defined on the full Hilbert space. The Schrödinger equation should be replaced with the unitary evolution

$$\psi(t) = \exp(-i\mathbf{H}t)\psi(0).$$

The Hamiltonian operator **H** is the infinitesimal generator of exp(-i**H**t), and is not necessarily bounded. When $\psi(t)$ is in the domain of $H$, the Schrödinger equation can be obtained as a special case of the unitary evolution according to Stone's theorem [9].

Adopting the unitary evolution exp(-i**H**t) offers an advantage over the Schrödinger equation: it is fully deterministic and is well defined for any time and initial state. Earman [6] takes this as the positive feature of quantum mechanics that avoids the singularities in classic Newtonian mechanics[1]. In contrast, the Schrödinger equation of a system may fail to apply under certain initial conditions after finite period of time. Thus, it is unnecessary, even undesirable, to restrict the state to Hamiltonian's domain.

Regarding the first difficulty, I argue that states with infinite expectation values of physical observables should not be considered as unphysical. In von Neumann's [17] standard formalism of quantum mechanics, expectation values of any physical quantities are never directly observed. Each individual observation yields a specific result, and the expectation value is only obtained as the average of long-term observations of an ensemble in the same quantum state. If $<X>_\psi$ is infinite, the average of long-term measurements of $X$ would not converge for an ensemble in the same state $\psi$, but this is not by itself "unphysical".

---

[1] As Earman has noticed, Newtonian mechanics is not fully deterministic. In systems of particles under Newtonian gravitational interaction, certain initial conditions lead particles to spatial infinity in finite time, rendering later evolution ill-defined. Quantum mechanics avoids this, provided the Hamiltonian is self-adjoint.

Moreover, from within the Hilbert space formalism of quantum mechanics, the probability distribution of each measurement result of $X$ can still be obtained when $\psi\notin\mathrm{Dom}(X)$ via the spectrum decomposition of $X$

$$X=\int\lambda_X \mathrm{d}P_{\lambda_X}$$

where $\lambda_X$ fall in $X$'s spectra and $\mathrm{d}P_{\lambda_X}$ is its corresponding projection-valued measure that satisfies

$$\int \mathrm{d}P_{\lambda_X}=1.$$

The probability that a single measurement result falls within a Borel measurable set $\varepsilon\in\mathbb{R}$ is simply

$$\int_\varepsilon \mathrm{d}P_{\lambda_X},$$

just as when $<X>_\psi$ is finite.

Therefore, falling outside the domains of unbounded operators in quantum mechanics does not hinder obtaining time evolution or measurement probability distributions. Unless we have compelling reasons to require that $<X>_\psi$ must be finite, states $\psi$ that make $<X>_\psi$ infinite should not be deemed unphysical. Yet Carcassi, Calderón, Aidala, and Heathcote only provide little justification for such requirement[2]. In addition, as I will discuss in the next section, it is untenable to require that $<X>_\psi$ must be finite for any self-adjoint operators $X$, and deciding which operators must have finite expectation values or not would rather intriguing.

## 3. Restricting states of the Hilbert space and the problem with the Schwarz space

Insisting on restricting the Hilbert space to exclude "unphysical" states immediately raises the challenge of selecting which operators $X$ require finite expectation values and which may allow infinite ones. Consider quantum mechanics on $\mathbb{R}$ for simplicity: as Carcassi, Calderón, and Aidala (p.18) have already noticed, no quantum states yield only finite expectation values for all operators $f(x)$ and $g(p)$, which are functions of position $x$ or momentum $p$. Moreover, only wave functions with compact support on $\mathbb{R}$ have finite expectation values for all operators in the form $f(x)$. This space of wave functions is not even closed under free evolution. In fact, if $\psi$ has compact support, $\exp(-i\mathbf{H_0}t)\psi$ does not have compact support for any $t \neq 0$ (where $\mathbf{H_0}$ denotes the free Hamiltonian operator) [12, p.64]! Therefore, the requirement that $<X>_\psi$ must be finite cannot be universal, and only a privileged subset of $X$ is selected to meet the requirement.

---

[2] An exception is that Heathcote [10, p.533] argues that states with infinite expectation values of the Hamiltonian need infinite energy to be prepared, which I will discuss in the next section.

Carcassi, Calderón, and Aidala choose this privileged subset of operators as polynomials of position and momentum spanned by operators in the form $x^n p^m$. Correspondingly, quantum states $\psi$ with finite expectation values of all polynomial operators form the Schwartz space $\mathcal{S}(\mathbb{R})$, a proper subspace of the Hilbert space $L^2(\mathbb{R})$. On the other hand, other operators like $e^x$ and $e^p$ may have infinite expectation values on states in $\mathcal{S}(\mathbb{R})$.

$\mathcal{S}(\mathbb{R})$ has many desirable mathematical features to be considered as a natural physical space according to Carcassi, Calderón, and Aidala. Nevertheless, I argue that the requirements that $<x^n p^m>_\psi$ must be finite is dubious. First, according to Carcassi, Calderón, and Aidala, $<x>_\psi$ must be finite while $<e^x>_\psi$ can be infinite. Yet they are measured by the same procedure, as any measurement of $x$ is by itself a measurement of $e^x$, *vice versa*. In a simplified setting of experiment, suppose that the apparatus yields a raw data $\xi_i$ for each measurement. Then, the values of $x$ and $e^x$ can be expressed as functions $A(\xi_i)$ and $B(\xi_i)$ respectively. The expectation values $<x>$ and $<e^x>$ are obtained approximately as $(A(\xi_1)+\ldots+A(\xi_n))/n$ and $(B(\xi_1)+\ldots+B(\xi_n))/n$. Unless the expectation $<x>$ is employed in some fundamental physical process, allowing non-convergence for the *B*-average while forbidding it for the *A*-average seems arbitrary.

Second, the convergence of $<x>_\psi$ relies on $\psi$'s behaviour at spatial infinity, and we have reasons to disregard its physical significance. When considering the behaviour of the wave function within a bounded spatial region that corresponds to a projection operator $P$, the wave function's expectation value of the position within the region, as calculated by $<x>_{P\psi}$, is always finite. It is unlikely that in successive more fundamental physical theory, the global expectation value $<x>_\psi$ has a crucial role in its structure.

Heathcote [10, p.533] argues that states with infinite expectation values of the Hamiltonian $\langle H \rangle_\psi$ are "physically impossible", as they need infinite amount of energy to be prepared. I take this as a reasonable concern. However, inability to be prepared in laboratories does not mean that these states cannot exist in nature. Moreover, if Heathcote's requirement is taken seriously, to decide the space of all physically possible states $S$, we must first decide the set of physically possible Hamiltonians $S_H$ that for any $H \in S_H$, $\langle H \rangle_\psi$ must be finite for any $\psi \in S$. As I will show later in this section, this requirement is not satisfied if we choose $S=\mathcal{S}(\mathbb{R})$ and include some reasonable Hamiltonians as physically possible.[3]

I have argued that there are no compelling reasons to restrict the Hilbert space as the space for physical states in quantum mechanics. In addition, I will further demonstrate that such restriction would raise intriguing problems regarding the relations between admissible physical states and admissible dynamic evolutions.

As already demonstrated, this space of functions with compact support on $\mathbb{R}$ is not closed under free Hamiltonian evolution. In general, stricter restrictions on the physical admissible states bring stricter restrictions on possible Hamiltonian evolutions. Conventional Hilbert space formalism of quantum mechanics poses no restriction on either physically admissible states or evolutions. On the other hand, restricting physical admissible states on $\mathcal{S}(\mathbb{R})$, as Carcassi, Calderón, and Aidala propose, would exclude a class of meaningful Hamiltonian evolutions, which is another undesirable feature of their suggestion.

For simplicity, I will only consider one-particle systems. In quantum mechanics, the Hamiltonian operator of a system is

$$H = p^2 + V(x)$$

where $p$ and $V(x)$ are the momentum and potential operators respectively. Strictly speaking, $H$ is usually unbounded and the aforementioned definition has to take into consideration the intricate relations between the domains of operators in the definition. In a more rigorous approach, an operator $p^2+V(x)$ on the space of infinitely differentiable functions with compact support $C_c^\infty(\mathbb{R})$ is defined. If it is essentially self-adjoint, its closure $H$ is an (unbounded) self-adjoint operator [6, 9].

---

[3] The best strategy, I think, is not to take the set of physically possible states and the set of physically possible Hamiltonians as *independent*. As I do not think it is necessary to restrict physically possible states in ordinary quantum mechanics, I defer this refinement to the last section.

Now, consider what restrictions $V(x)$ should have to ensure that $\mathcal{S}(\mathbb{R})$ is closed under the unitary evolution exp(-i$H$t). According to the Ehrenfest theorem, the time derivate of the expectation value of an operator $A$ satisfies
$$d<A>_\psi/dt=i<[H,A]>_\psi.$$
If $V(x)$ is a polynomial of $x$, then for any polynomial $A$ of position and momentum, $[H,A]$ is also a such polynomial. Consequently, d$<A>_\psi$/d$t$ is finite, and it follows that exp(-i$H$t)$\psi\in\mathcal{S}(\mathbb{R})$.

However, the closeness of $\mathcal{S}(\mathbb{R})$ breaks down when $V(x)$ is not a polynomial of $x$. For some $V(x)$ with negative powers of $x$, $p^2+V(x)$ is still essentially self-adjoint [6], including physically meaningful examples, such as the Coulomb potential. We have
$$d<p>_\psi/dt=<-\partial V/\partial x>_\psi.$$
Since $\partial V/\partial$x is not a polynomial of $x$, for some $\psi\in\mathcal{S}(\mathbb{R})$, d$<p>_\psi$/d$t$ may diverge, and exp(-i$H$t)$\psi$ may depart the Schwartz space $\mathcal{S}(\mathbb{R})$. Polynomials of the position cover a large variety of interactions in physics, but excluding negative-power potentials is too great a sacrifice.

## 4. Further problems regarding the dynamics

In critiquing Carcassi, Calderón, and Aidala's proposal, Lemos [11] worries that self-adjointness of operators cannot be defined on $\mathcal{S}(\mathbb{R})$. Therefore, it is questionable whether the spectral theorem—crucial for quantum dynamics—can be reconstructed. Lemos questions whether physical observables can be defined as intrinsically essentially self-adjoint operators on $\mathcal{S}(\mathbb{R})$ without invoking Hilbert space structure:

> It is hard to envisage how the notion of essential self-adjointness can be defined without referring to some extension of the operator's domain, which would inevitably involve going beyond Schwartz space. (p.8)

Even if this can be done, Lemos continues,

> One of the basic tenets of quantum mechanics could be rephrased to state that to each measurable quantity there corresponds an essentially self-adjoint operator. It remains to be seen whether this is a fruitful line of inquiry. (*ibid.*)

Heathcote [10, p.529] similarly worries that as many physical operators are no longer self-adjoint when restricted to narrow domains, challenging the fundamental postulate in quantum mechanics that observables are represented by self-adjoint operators.

In their pioneering work on axiomatic quantum field theory, Streater and Wightman [16] has already noticed the great difficulties brought by restricting the domains of operators:

> Having once resigned ourselves to dealing with unbounded operators, we face a couple of practical problems having to do with their domains. … One is led naturally to assume a common dense domain, $D$, for all the unbounded operators in question … Once one has $D$, another problem arises: to what extent do the values of the unbounded operators on $D$ determine them uniquely wherever else they can be defined? This is particularly acute for observables because an operator which is hermitian[4], when restricted to vectors in $D$, might have several different self-adjoint extensions, and to specify a theory one would have to tell which self-adjoint extension is meant. (pp.90-91)

In this section, I will explicate their worry.

There does not exist a one-to-one correspondence between essentially self-adjoint operators on $\mathcal{S}(\mathbb{R})$ and self-adjoint operators on $L^2(\mathbb{R})$. A self-adjoint operator's restriction on $\mathcal{S}(\mathbb{R})$ must be symmetric but may fail to be essentially self-adjoint. Conversely, any essentially self-adjoint operator on $\mathcal{S}(\mathbb{R})$ uniquely determines a self-adjoint operator on $L^2(\mathbb{R})$. Therefore,

---

[4] In other words, symmetric.

~ 9 ~

the essentially self-adjoint operators on $\mathcal{S}(\mathbb{R})$ determine a proper subset of self-adjoint operators on $L^2(\mathbb{R})$, and we have reasons to believe that all physically "realistic" Hamiltonians belong to this subset. As shown in the last section, most Hamiltonians are defined as the closures of essentially self-adjoint operators $p^2+V(x)$ on $C_c^\infty(\mathbb{R})$. These essentially self-adjoint operators all correspond to essentially self-adjoint operators on $\mathcal{S}(\mathbb{R})$, since $C_c^\infty(\mathbb{R})\subset\mathcal{S}(\mathbb{R})$.

Other self-adjoint operators on $L^2(\mathbb{R})$ are conceivable as the Hamiltonians of a system. However, they are not uniquely determined by classical Hamiltonians and would break the classical-quantum correspondence. These Hamiltonians are usually too exotic and are usually excluded from the considerations in physics. For example, $H_1=p^2-x^4$ defined on $\mathcal{S}(\mathbb{R})$ is symmetric but not essentially self-adjoint [9]. In general, a symmetric but non essentially self-adjoint operator may admit no self-adjoint extensions or infinite self-adjoint extensions [8]. $H_1$ has infinite self-adjoint extensions and none is "canonical": There exists a state $\varphi\in L^2(\mathbb{R})$ such that $(\varphi,(p^2-x^4)\varphi)\neq((p^2-x^4)\varphi,\varphi)$ [9], and therefore any self-adjoint extension of $H_1$ cannot preserve the form of $p^2-x^4$ on certain states even if $p^2-x^4$ can be defined on these states. $H_1$'s classical Hamiltonian counterpart has the exotic feature that particles can be expulsed to spatial infinity within finite period of time. It should be expected that realistic physical systems should not have increasing or non-diminishing repulsive force at spatial infinity. In fact, for most realistic physical systems, their quantum Hamiltonians, when restricted on $C_c^\infty(\mathbb{R})$, are essentially self-adjoint [13]. If $\mathcal{S}(\mathbb{R})$ is closed under these Hamiltonian revolutions as discussed in the last section, one can continue to use the Stone theorem and spectral decomposition for quantum mechanics on $\mathcal{S}(\mathbb{R})$.

Therefore, Heathcote's worry may be alleviated, but what is taken as additional "realistic" restrictions in axiomatic ordinary quantum mechanics is now shifted as the fundamental mathematical structure for quantum mechanics on the Schwartz spaces. This difference will be further discussed in the next section.

Now turn to Lemos's worry. The definition of essentially self-adjointness, naïvely, relies on the structure of Hilbert space larger than the Schwartz spaces. Most criteria of essentially self-adjointness also involve operators defined on space larger than the Schwartz spaces (for example, the adjoint of symmetric operators on the Schwartz spaces). Yet, on the other hand, any Schwartz space uniquely determines a Hilbert space as its completion with respect to its norm. I take it indisputable that the norm is an "intrinsic" property of a Schwartz space. Then, features resorting to the structure of $L^2(\mathbb{R})=\mathcal{S}(\mathbb{R})^*$ can be understood as "intrinsic" features within $\mathcal{S}(\mathbb{R})$ in an indirect way, and I suspect whether it is possible to provide a clear account to classify the "intrinsic" features of a mathematical structure such as $\mathcal{S}(\mathbb{R})$. Can "belonging to $(\pi-1, \pi+1)$" be understood as properties relying only on intrinsic features of $\mathbb{Q}$? I suspect whether pursuing a such account is necessary for doing physics.

I suggest that our attitudes towards Lemos's worry depends on how we interpret Carcassi, Calderón, and Aidala's proposal. In a more radical interpretation, they replace the received axiomatic quantum mechanics with a new formulation with a set of axioms defined on the Schwartz spaces. Then, the simplicity of these axioms would be questioned if they too often rely on mathematical structures larger than the Schwartz spaces. In a more conservative interpretation, they do not replace the axioms of ordinary quantum mechanics on Hilbert spaces. Instead, they divide the states according to the formalism of quantum mechanics into two classes: the "physical" and the "unphysical" states. In this interpretation, we do not redefine quantum Hamiltonians of systems as essentially self-adjoint, and can continue to use any self-adjoint operators of the Hilbert spaces.

**5. The vagueness of "physicality"**

~ 11 ~Schwartz spaces. Most criteria of essentially self-adjointness also involve operators defined on space larger than the Schwartz spaces (for example, the adjoint of symmetric operators on the Schwartz spaces). Yet, on the other hand, any Schwartz space uniquely determines a Hilbert space as its completion with respect to its norm. I take it indisputable that the norm is an "intrinsic" property of a Schwartz space. Then, features resorting to the structure of $L^2(\mathbb{R})=\mathcal{S}(\mathbb{R})^*$ can be understood as "intrinsic" features within $\mathcal{S}(\mathbb{R})$ in an indirect way, and I suspect whether it is possible to provide a clear account to classify the "intrinsic" features of a mathematical structure such as $\mathcal{S}(\mathbb{R})$. Can "belonging to $(\pi-1, \pi+1)$" be understood as properties relying only on intrinsic features of $\mathbb{Q}$? I suspect whether pursuing a such account is necessary for doing physics.

I suggest that our attitudes towards Lemos's worry depends on how we interpret Carcassi, Calderón, and Aidala's proposal. In a more radical interpretation, they replace the received axiomatic quantum mechanics with a new formulation with a set of axioms defined on the Schwartz spaces. Then, the simplicity of these axioms would be questioned if they too often rely on mathematical structures larger than the Schwartz spaces. In a more conservative interpretation, they do not replace the axioms of ordinary quantum mechanics on Hilbert spaces. Instead, they divide the states according to the formalism of quantum mechanics into two classes: the "physical" and the "unphysical" states. In this interpretation, we do not redefine quantum Hamiltonians of systems as essentially self-adjoint, and can continue to use any self-adjoint operators of the Hilbert spaces.

**5. The vagueness of "physicality"**

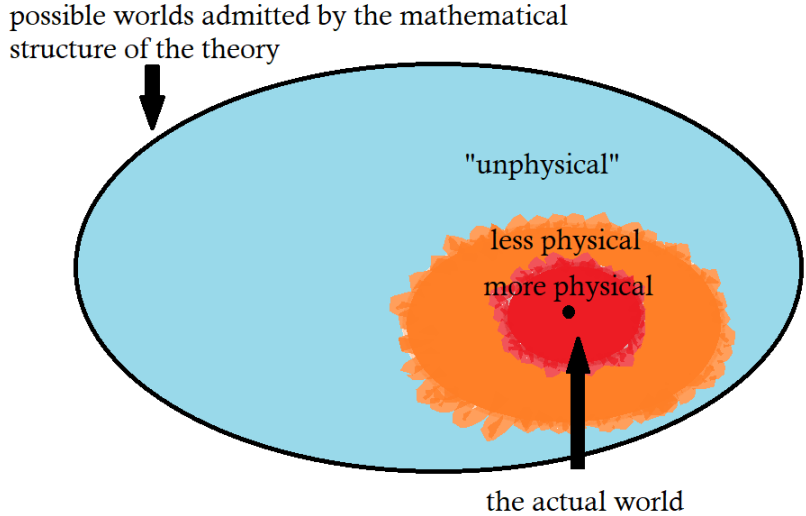

Fig. 1: Different levels of "physical" possible worlds

This paper serves as an exposition on the "physicality" or "unphysicality" of certain states in the Hilbert spaces in quantum mechanics, but no definition of "physicality" has been provided so far. Intuitively, it relates to possible worlds. The more physical a world is, the more likely it qualifies as possible relative to our actual world. Conversely, the more unphysical, the more likely it is impossible. However, "being physical" may admit multiple levels of degree, which corresponds to a hierarchy of possible worlds. First, I will consider the smallest and the biggest class of "physically possible worlds":

(Hierarchy I) Our world has its unique history and dynamics. Our world (or history) is the only possible world (or history). In other words, the set of possible worlds only admit one member. This view is also called "strong determinism" [3, 4], according to which our actual world is the only physical world.

(Hierarchy II) Physical possible worlds are precisely the models of fundamental physics' mathematical structure. The boundary between physical and unphysical worlds is clear. For example, a spacetime represented by a tripe $(\mathcal{M}, g, T)^5$ is physical if and only if it satisfies Einstein's field equation.

---

[5] $\mathcal{M}$, $g$, $T$ stand for the manifold of spacetime, its metric, and its energy-momenta tension field.
~ 12 ~

These two levels admit no vagueness, as shown in Fig. 1. However, there are also many intermediate layers of possibility between them, as we identify some possible worlds with some important "realistic" features with the actual world while others do not. A third class of hierarchies can be introduced:

> (Hierarchy III) The physical possible worlds are the models of the mathematical structure in fundamental physics that also meet certain "realistic" requirements.[6]

As shown in Fig. 1, this hierarchy is ambiguous, as there is no unequivocal way to flesh out the requirements of being "realistic". Furthermore, this level comprises sub-layers, as the importance and necessity of "realistic" can admit degrees. For example, if we demand determinism, a sub-level can be introduced

> (Hierarchy III.1) The physical possible worlds are the models M of the mathematical structure in fundamental physics that also meet deterministic requirements. Loosely speaking, a model of classical mechanics on $\mathbb{R}^n$ satisfies deterministic requirements if the potential satisfies the Lipschitz condition. A model of quantum mechanics satisfies deterministic requirements if its Hamiltonian operator is self-adjoint. A spacetime represented by tripe ($\mathcal{M}, g, T$) is physical if it is global hyperbolic[7].

Unlike classical mechanics and general relativity, the set of physical possible worlds according to (Hierarchy III.1) coincide with the set of physical possible worlds according to (Hierarchy II).

Moreover, Carcassi, Calderón, and Aidala's proposal requires a more

---

[6] The "past hypothesis" that the initial state of the universe must have low entropy as discussed by Chen [3] also falls in this category in my view, as it is a restriction of possible states in a given mathematical formalism.

[7] I do not want to put a fully rigorous and comprehensive characterisation on the determinism in general relativity. For more discussions, see [15].

nuanced criterion of being "realistic":

> (Hierarchy III.2) The physical possible worlds are the models M of the mathematical structure in fundamental physics. Besides, their states and dynamics must meet certain restrictions. For example, in general relativity, the models must satisfy certain energy conditions [8]. In quantum mechanics, certain observables must have finite expectation values.

I have demonstrated in this paper that adopting the criterion of physicality according to (Hierarchy III.2) in quantum mechanics is unpromising and lacks motivations. Furthermore, it is also difficult to obtain a consistent rigorous mathematical expression of this criterion in quantum mechanics. It is difficult to flesh out which observables must have finite expectation values and which may without arbitrariness. Posing such restrictions would also simultaneously limit possible forms of Hamiltonian operators, and the restrictions cannot be too loose (including $p^2$-$x^4$ as "physical") nor too strict (excluding physically meaningful ones like $p^2$-$1/x$).

    Furthermore, as noticed in the last section, (Hierarchy II) and (Hierarchy III) may interconvert. If we can successfully build a rigorous mathematical formalism of quantum mechanics on mathematical spaces smaller than the Hilbert spaces, what is taken as additional "realistic" requirements according to (Hierarchy III.2) in quantum mechanics on Hilbert spaces are now transformed as the requirements following the mathematical structures of theories as in (Hierarchy II). This shows that the boundary between physically possible worlds and impossible worlds is inevitably vague, and it should be caveated what "physicality" one bears in mind in discussions.

## 6. Concluding remarks: semiclassical quantum gravity

Carcassi, Calderón, and Aidala's proposal arises from concerns over infinite expectation values for certain observables. In standard quantum mechanics, as I have argued, there are no compelling reasons to avoid such infinities, but similar concerns may be justified in other fundamental theories. In semiclassical quantum gravity, the expectation values of energy-momenta field operators appear in its fundamental equation: the semiclassical Einstein equation [18, p.86]

$$G=8\pi <\bm{T}>$$

where G is the Einstein tensor field and $\bm{T}$ is the energy-momenta operator-valued field, analogous to the classical Einstein equation

$$G=8\pi T.$$

---

[8] For more on the energy conditions, see [5].

To avoid spacetime singularities, the equation's right-hand side must remain finite[9]. Therefore, some physicists require that all possible physical states in quantum field theory must satisfy the so-called Hadamard condition [7]. This restriction stems from mathematical consistency in further theoretical constructions, instead of merely being "realistic" relative to the actual world. As Ruetsche [14, p.239] has noticed, the Unruh vacuum, which is physically significant in understanding Hawking radiation and black hole thermodynamics, does not satisfy the Hadamard condition. Ruetsche thus suggests that the physicality of non-Hadamard states depends on explanatory or developmental motivations.

    Quantum field theory and the Hadamard condition also provide a natural setting where the physicality of quantum states and the physicality of Hamiltonian revolutions should not be taken as independent. If we adopt the algebraic quantum field theory approach, a quantum state is defined as a functional $\omega$ of a C* algebra $\mathcal{U}$ representing the quantum observables of a system. The dynamic evolution of the system is represented as a one-parameter group of automorphisms $\{\alpha_t\}$ of $\mathcal{U}$. The energy-momenta operator-valued field $\bm{T}$ at different spacetime points rely on the group $\{\alpha_t\}$. Therefore, the interdependence of quantum states and dynamic evolutions is imprinted in the very definition of the Hadamard condition. Semiclassical quantum gravity presents a nuanced challenge in understanding the physicality of mathematical structures in physics and the hierarchies of "physicality" outlined previously. Therefore, I invite more research from the physicality of Hilbert spaces in quantum

---

[9] Strictly speaking, after regularization.

mechanics to the physicality of structures in quantum field theory.

**Conflict of Interest Statement:** The author states that there is no conflict of interest.

**Data Availability Statement:** The author does not analyse or generate any datasets, because the work proceeds within a theoretical and mathematical approach